**Connection between Phantom and Spatial Correlation in the Kolmogorov-Johnson-Mehl-Avrami-model: A brief review.**

By

M. Tomellini [a] and M. Fanfoni [b]

a) Dipartimento di Scienze e Tecnologie Chimiche, Università di Roma Tor Vergata, Via della Ricerca Scientifica, 00133, Roma, Italy

b) Dipartimento di Fisica, Università di Roma Tor Vergata, Via della Ricerca Scientifica, 00133, Roma, Italy

**Abstract**

The goal of this minireview is restricted to describe how the Kolmogorov-Johnson-Mehl-Avrami model has evolved from its birth up to the present day. The model, which dates back to the late of 1930s, has the purpose of describing the kinetics of a phase transformation. Given the nature of this article, although there are hundreds (if not thousands) of experimental data concerning the most disparate topics, which are interpreted on the basis of the KJMA model, no arguments relating to these, will be touched upon.

Starting from the ingenious concept of phantom nuclei, firstly introduced by Avrami to get the exact kinetics, we review theoretical approaches which overcome such concept. We will show how spatial correlation among nuclei plays a fundamental role in these modelings.

**1-Introduction and problem description.**

A Poisson point process in any space dimension D, D being 1, 2 or 3, implies the possibility to place in a completely random way, for instance sequentially, a certain number of points, say $N_0$, in a given "volume". By definition, the probability that two points may exist at very close distance is different from zero. Conversely, if points are precluded from lying at a distance less than, say $R$, from each other, their spatial distribution will be not completely random, in fact the imposed constraint introduces a certain degree of correlation among them. It goes without saying that correlation implies the existence of an interaction potential between the points which, in the case just described, is an infinite barrier far $R$ from each point (Fig.1).

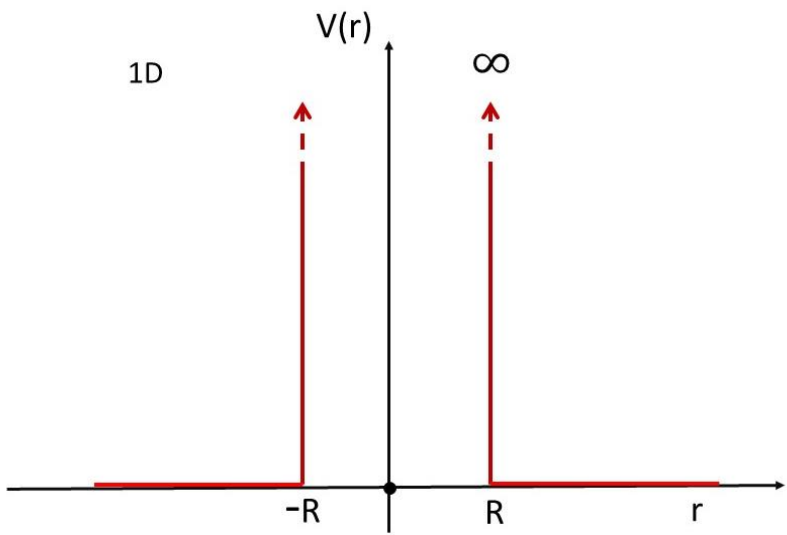

Fig.1. An example of hard-core pair interaction potential in 1D.

Think now of a process in which, at time $t$, from points of a certain D-space, D-spheres begin to grow with their radii following a deterministic law $R(t)$. This process simulates a phase transformation that proceeds by nucleation and growth.

We can distinguish two cases:

1) the D-sphere centers (*dots*) arise distributed at random and start growing simultaneously. This process will be referred as: Dirac delta nucleation.

2) the birth of the D-sphere centers is not simultaneous and is described by a deterministic time law.

A.N. Kolmogorov (1937), W.A. Johnson and R. F. Mehl (1939) and M. Avrami (1939-1941) (KJMA), have independently, faced up to and solved the two cases, in the sense that they were able to give the analytic expression of the kinetics of coverage [1-5]. It is worth pointing out that the three contributions provide the solution by means of different approaches: respectively probability theory (K), rate equations (JM) and set theory (A). It is also enlightening the discussion between Johnson and Mehl and Avrami on the topic reported in the paragraph "DISCUSSION" of ref.[5]. In fact, it was centered on the nucleation events taking place in the already transformed phase, namely what later Avrami will refer to as phantom nuclei.

Since then, the model has been extensively employed for describing phase transformations by nucleation and growth in chemistry [6], materials science [7-11] and electrochemistry [12,13]. Furthermore, some theoretical aspects of the model are relevant also in the areas of mathematics [14,15], biology [16,17,18,19], medicine [20,21] and sociology [22].

The possibility to apply the model to describe real systems has been thoroughly discussed in literature by studying the effects on kinetics of anisotropic growth [23,24,25], nucleation at defects and at grain boundaries [2,26,27,28], phase transformations in different metrics [29], phase change under non-isothermal conditions [30,31] and with non-random nucleation [32-34]

Let us detail a little further the case 2). The formation of new nuclei can take place at random only in the space that has not been covered by the other growing D-spheres. In the ensuing, we refer to the D-sphere with the noun nucleus and to its center as dot. It is clear, then, that this condition leads back to the aforementioned correlated nucleation. The border of the transformed phase plays the role of the infinite barrier potential mentioned above, consequently the centers of the D-spheres are correlated.

Summing up: the substantial difference between the cases 1) and 2) is that the first, the Dirac delta nucleation, is a genuine Poisson process i.e. the centers from which the new phase starts forming are randomly dispersed throughout the entire D-volume, whereas the second is a correlated process.

It is important to underline that, beyond any other effect such as, for instance, elastic interaction or diffusion, the simple time dependence of the nucleation is enough to determine a correlation condition. Evidently, once one finds a way to approach the problem of correlated nuclei with a simple interaction potential, the same calculation technique can be applied to more complex potentials.

This paper, which follows a previous brief review of us where we discussed the original KJMA model based on genuine Poisson process [4], is organized as:

In section two, the concept of phantom in KJMA theory and its limits are described. Section three is devoted to the series which allow to determine the transformed volume. Basically, they are three: Avrami series, based on the set theory and its rewriting in two forms: the first, in terms of the distribution functions, $f_k(\boldsymbol{r}_1, \ldots, \boldsymbol{r}_k)$, the second, in terms of the correlation functions $g_k(\boldsymbol{r}_1, \ldots, \boldsymbol{r}_k)$[1]. These series establish the link with the Statistical Mechanics and allows to get rid of phantoms. In section four we report numerical computations on KJMA compliant transformations in terms of actual nuclei, only. In the same section the question of phantom overgrowth is reviewed. After the last section, devoted to the conclusions, we have included an Appendix where detailed calculations of some formulae of the main text have been reported in order to make reading smoother and easier.

## 2-The concept of phantom.

With reference to case 1) above, the kinetics of transformation can be obtained, easily, using the Poisson distribution, $p_n(m) = \frac{m^n}{n!} e^{-m}$, that gives the probability that $n$ events in space or time, occur in a given domain, $m$ being the mean value of $n$ in that domain. The KJMA formula (case 1) for nucleus radius $R$, is obtained through $p_0$. In fact, the fraction of untransformed phase[2], $1 - V$, is equal to the probability that a generic point, say c, of the space does not belong to any D-spheres, i.e. $V = 1 - p_0$. This requirement implies that, given a region of radius $R$ encompassing the point c, this region is empty of nuclei centers. Therefore, $p_0 = e^{-m} = e^{-N\omega_D R^D}$, where $N$ is the number of nuclei per unit volume, and $\omega_D R^D$ is the volume of the D-sphere with $\omega_1 = 2$; $\omega_2 = \pi$; $\omega_3 = \frac{4}{3}\pi$ and then

$$V = 1 - e^{-N\omega_D R^D}. \qquad (1)$$

It goes without saying that being $R = R(t)$, eqn.1 gives the kinetics $V(t)$.

---

[1] For the notation of *f* and *g* functions see the note (3).

[2] From now on volumes must be considered as normalized to the entire volume of the system. The transformation is completed when $V = 1$.

As far as the case 2) is concerned, in contrast to the case 1) it is not a genuine Poisson process. Nevertheless, it can be led back to a Poisson process by considering a random nucleation to take place everywhere in the entire volume system. However, due to non-simultaneous nucleation, it happen that a nucleus may born in a region of the space already transformed: such a nucleus is called a *phantom* [3]. It does not contribute to the actual transformed phase but cannot be avoided in the mathematical formulation to get the correct solution. A discrete set of $N_i$ nucleation events (per unit volume) at time $t_i$ ($i = 1, 2, ...$), give rise to as many nuclei of radius $R_i$ ($i = 1, 2, ..., $). However, a subset of these can be born within nuclei of radius $R_j$ born at time $t_j < t_i$: this is a subset of phantoms. Moreover due to the independence of the events, $p_0 = \prod_i p_{0,i} = \prod_i e^{-N_i \omega_D R_i^D} = e^{-\sum_i N_i \omega_D R_i^D}$. Therefore, $N_i$ comprehends phantom nuclei. The continuum limit of this equation provides

$$V(t) = 1 - e^{-\omega_D \int_0^t \frac{dN}{d\tau} R^D(t-\tau)d\tau} \quad , \qquad (2)$$

where $\frac{dN}{d\tau}$ is the nucleation rate and $\tau$ is the "birth" time of nuclei. Eqn.2 is the celebrated KJMA formula. We stress that the use of Poisson process imposes a constraint on the growth law of nuclei, since a growing phantom must not overtake a real nucleus. Specifically, KJMA compliant growth law are either linear or convex $R(t)$ functions ($\frac{d^2R}{dt^2} \geq 0$). On the contrary, concave functions ($\frac{d^2R}{dt^2} < 0$) give rise to phantom overgrowth [35,36]. Fig.2 illustrates the case of linear growth: $R(t) = \mathrm{v}t$.

We find that the elegance of the KJMA theory lies on the paradox that in order to obtain the *exact* mathematical solution, non-real "entities" (phantoms) must be considered.

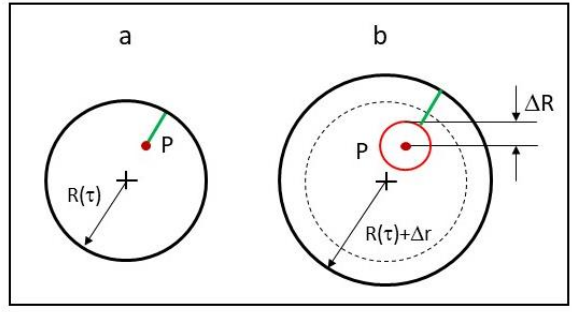

Fig.2. In the linear growth $\Delta R = \text{v}\Delta\tau$ where v is the growth rate. Panel a): Actual nucleus that was born at time equal zero, at running time $\tau$. The red dot is a phantom that starts growing at time $\tau$ . Panel b) shows the same system as in a) at $\tau + \Delta\tau$ and $R + \Delta R$. Notably, since the lengths of the green segments are equal, i.e. it does not change during the growth, it follows that phantom overgrowth is precluded.

## 3-Series

The approach which makes use of actual nuclei needs considering correlated nucleation. In this section, we present the theoretical formulation for studying phase transformations with correlated nuclei.

### 3.1 Avrami series

In the paper "Kinetics of phase change I: General theory" [3] Avrami exploited the set theory to estimate the volume occupied by an ensemble of overlapping D-spheres, in general not-equal in size. This is the morphology of the transformed phase ruled by impingement mechanism of growing nuclei. Avrami derived an interesting series which converge to the measure of the transformed phase $V(t)$, it is

$$V = V_{1,ex} - V_{2,ex} + \cdots (-)^{k+1}V_{k,ex} + \cdots = \sum_{k=1}^{\infty} (-)^{k+1}V_{k,ex} \quad (3)$$

$V_{k,ex}$ refers to the whole volume occupied by at least $k$ overlapping spheres. The meaning of $V_{k,ex}$ will be made clearer through an example. In the following, we also denote with $v_k$ the volume of the overlap region of *exactly $k$-spheres*. Here we give a demonstration of eqn.3 for the 2D case, the generalization to any D is straightforward; furthermore, we limit ourselves to the case of four overlaps being the cases with number of overlap greater than four conceptually the same, only more cumbersome. The structure of the four overlapping nuclei has been thought for simplifying the description, in fact the nuclei are positioned in a way that can be referred as North, East, South, West without misunderstanding.

The cluster of four overlapping nuclei is reported in Fig.3 where the color code evidences the number of overlaps. In the figure, the extended volumes are also evidenced separately.

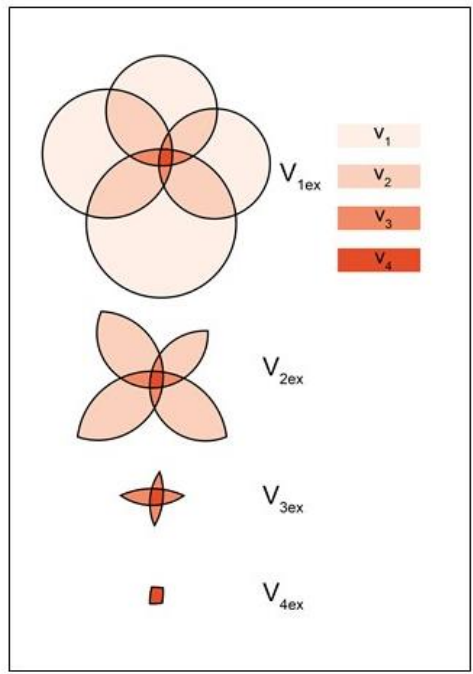

Fig.3. Four overlapped nuclei where the color code evidences the number of overlaps. The extended volumes are also displayed separately. For the detailed analysis of extended volumes see the text.

It is quite easy to verify that

$$V_{1,ex} = v_1 + 2v_2 + 3v_3 + 4v_4 = \binom{1}{1} v_1 + \binom{2}{1}v_2 + \binom{3}{1}v_3 + \binom{4}{1}v_4. \qquad (4a)$$

As far as the $V_{2,ex}$ is concerned, the extended volume of a single overlap (in the figure they are four), the volume $v_2$, naturally contribute once (i.e. $\binom{2}{2}$) to $V_{2,ex}$, whereas volumes $v_3$ and $v_4$ give three (i.e. $\binom{3}{2}$) and six (i.e. $\binom{4}{2}$) contributions, respectively; Fig.4a, Fig.4b, display graphically these contributions.

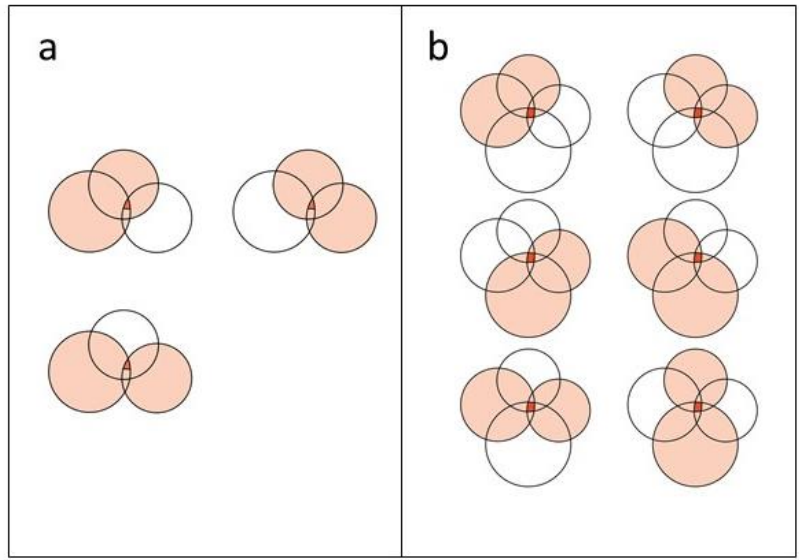

Fig.4. The contributions of $v_3$ and $v_4$ to the extended volume $V_{2,ex}$, are displayed in panel a and b, respectively. For the color code see Fig.3. In panel a only the contribution of the North pseudo-triangle is reported.

In particular, so as to describe the contribution of $v_3$ to $V_{2,ex}$, we decided to show only the case of the pseudo-triangle positioned at North. To this, three pairs of nuclei contribute i.e. EN-EW-NW, the South nucleus does not give any contribution. For the East, South and West pseudo-triangles the design is similar and, for each of them, there are always three contributions. The graphical explanation of the $v_4$ contribution is straightforward (see Fig.4b). Mathematically $V_{2,ex}$ is written as

$$V_{2,ex} = \; v_2 + 3v_3 + 6v_4 \; = \binom{2}{2}v_2 + \binom{3}{2}v_3 + \binom{4}{2}v_4. \quad (4b)$$

As regards the contributions to $V_{3,ex}$, obviously $v_3$ contributes once, whereas the volume $v_4$ contributes four times (i.e. $\binom{4}{3}$) as displayed in Fig.5.

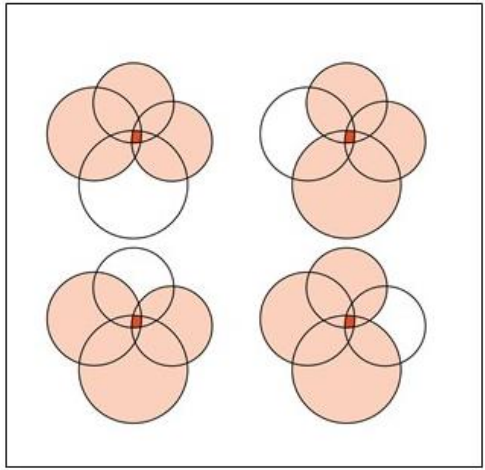

Fig.5. Contribution of $v_4$ to the $V_{3,ex}$. For the color code see Fig.3.

The following equation can be written:

$$V_{3,ex} = v_3 + 4v_4 = \binom{3}{3}v_3 + \binom{4}{3}v_4 \qquad (4c)$$

and finally

$$V_{4,ex} = \binom{4}{4}v_4 . \qquad (4d)$$

From Fig.3 it is easy to verify that

$$V = v_1 + v_2 + v_3 + v_4 \qquad (4e)$$

and exploiting eqns.4a-d it can be rewritten as

$$V = V_{1,ex} - V_{2,ex} + V_{3,ex} - V_{4,ex} \ , \qquad (4f)$$

that is eqn.3.

Before concluding this section we generalize eqns.4

$$V = \sum_{n=1}^{\infty} v_n \qquad\qquad (5a)$$

and

$$V_{k,ex} = \sum_{n=k}^{\infty} \binom{n}{k} v_n \ . \qquad\qquad (6a)$$

We note that eqn.3, determined by only taking into account the actual nuclei, was derived by Avrami and it holds for any spatial distribution of nuclei, whether it is random or not. For this reason, Avrami's work provides a more general contribution to the topic of phase change, as it laid the foundation for dealing with spatially correlated nuclei.

### 3.2. f -series[3]

The eqns.5a and 6a can be rewritten to highlight the link between Avrami's series and the probabilistic approach by Reiss et al [38], although eqns.3-6 hold for any distribution of sphere size. In particular, from eqn.5a

---

[3] In order to avoid any possible misunderstanding, a clarification is needed on this point. In the present article we use the notation employed by Van Kampen [37] where the $f_n$-functions are defined according to eqn.7. In Reiss et al paper Van Kampen 's $f_n$-functions are written as $g_n$ and named correlation functions. On the other hand, Van Kampen's correlation functions are defined through eqn.11 (section 3.3) and indicated as $g_n$.

$$V = \sum_{n=1}^{\infty} v_n = \sum_{n=1}^{\infty} p_n = 1 - p_0, \qquad (5b)$$

and from eqn.6a

$$V_{k,ex} = \sum_{n=k}^{\infty} \binom{n}{k} v_n = \sum_{n=k}^{\infty} \binom{n}{k} p_n , \qquad (6b)$$

where $p_n = v_n$ $(n \neq 0)$ *is the probability that a generic point of the space belongs to the region of overlap of exactly n nuclei* (Fig.3), ergo, $p_0$ *is the probability that the generic point belongs to the untransformed phase.*

The first case we want to discuss is a collection of D-spheres equal in sizes (which stems from Dirac's delta nucleation). In this case the definition of $p_n$ above, is equivalent to the following: $p_n$ *is the probability that the centers of exactly n nuclei lie in the spherical region $\Delta_R$ of radius R.* To be clearer, the probability that a generic point of the space belong to the overlap region of $n$-nuclei of radius $R$ is equivalent to the probability that a circle of radius $R$, centered at a generic point inside the overlap region, contains the centers of the $n$-nuclei. Panel a) of Fig.6, shows two nuclei whose centers are 1 and 2 which give rise to two overlapping circles, the overlap region is highlighted in green. A circle of radius $R$ centered at any point within the green region contains the points 1 and 2. It follows that the probability that two points fall within the dashed circle centered at A is equivalent to the probability that a point (for example point A in figure) belongs to the region of overlap of the two nuclei. Analogous reasoning for the case of three overlapping nuclei (purple region) leads to the representation displayed in panel b). It goes without saying that the same reasoning can be extended to $k$-nuclei.

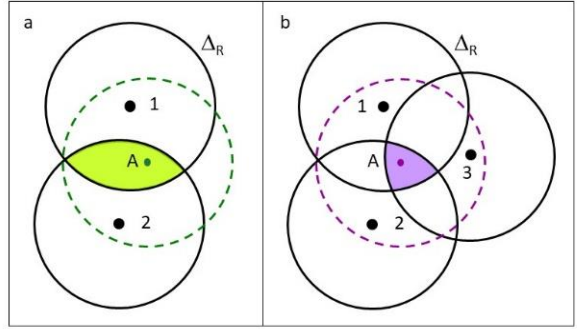

Fig.6. Simultaneous nucleation. Panel a): the probability a generic point of the space (point $A$ in the figure) belongs to the overlapping region ($v_2$) of two nuclei of radius $R$, is equal to the probability ($p_2$) that the centers of *exactly* two nuclei fall within the $\Delta_R$ domain, namely the circle of radius $R$ centered at $A$ (dashed line). Panel b): the probability a generic point of the space (point $A$ in the figure) belongs to the overlapping region ($v_3$) of three nuclei of radius $R$, is equal to the probability ($p_3$) that the centers of *exactly* three nuclei fall within the $\Delta_R$ domain, namely the circle of radius $R$ centered at $A$ (dashed line).

For $k = 2$ eqn.6b becomes

$$V_{2,ex} = 1p_2 + 3p_3 + 6p_4 + 10p_5 \dots . \qquad (6c)$$

The first three terms of this equation are graphically illustrated in Fig.7. Owing to the binomial coefficients in eqn.6c, $V_{2,ex}$ is equal to the average number of pair of nuclei within $\Delta_R$. Consequently, *$V_{k,ex}$ is equal to the average number of k-tuple within the $\Delta_R$ domain, that in Statistical Mechanics can be written by using distribution functions* :

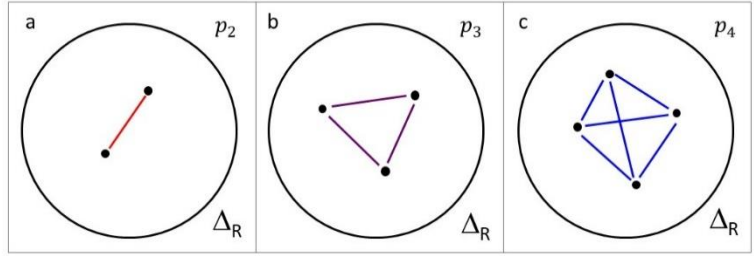

Fig.7. Graphical representation of $V_{2,ex}$ (eqn.6c). Panels a), b) and c) show the events related to $p_2$, $p_3$ and $p_4$, that exactly 2, 3 and 4 nuclei fall within $\Delta_R$. The binomial coefficients in eqn.6c give the number of couples for $n = 2$, $n = 3$ and $n = 4$, namely the number of segments in each panel.

$$V_{k,ex} = \frac{N^k}{k!} \int_{\Delta_R} f_k(\boldsymbol{r}_1, \dots, \boldsymbol{r}_k) d\boldsymbol{r}_1 \dots d\boldsymbol{r}_k, \qquad (7)$$

where $N$ is the number of nuclei per unit volume and $\frac{1}{V_0^k} f_k(\boldsymbol{r}_1, \dots, \boldsymbol{r}_k) d\boldsymbol{r}_1 \dots d\boldsymbol{r}_k$ is the probability of finding $k$ specific nuclei within volume elements $d\boldsymbol{r}_1, \dots, d\boldsymbol{r}_k$ at $\boldsymbol{r}_1, \dots, \boldsymbol{r}_k$, independently of the location of the other nuclei, with $V_0$ the whole volume where the transformation occurs [39,40]. The factorial term in eqn.7 corrects for the equivalent configurations obtained by permutation of $\boldsymbol{r}_1, \dots, \boldsymbol{r}_k$ coordinates. As an example, let us illustrate the case $k = 2$. The probability of having a specific couple ( 1,2 ) within $\Delta_R$ is $\frac{1}{V_0^2} \int_{\Delta_R} f_2(\boldsymbol{r}_1, \boldsymbol{r}_2) d\boldsymbol{r}_1 d\boldsymbol{r}_2$. To determine the mean number of couples in $\Delta_R$, i.e. $V_{2,ex}$, we have to multiply this probability by the total number of possible couples, namely $\frac{M(M-1)}{2} \cong \frac{M^2}{2!}$ where $M = NV_0$ is the total number of nuclei, to obtain $V_{2,ex}$.

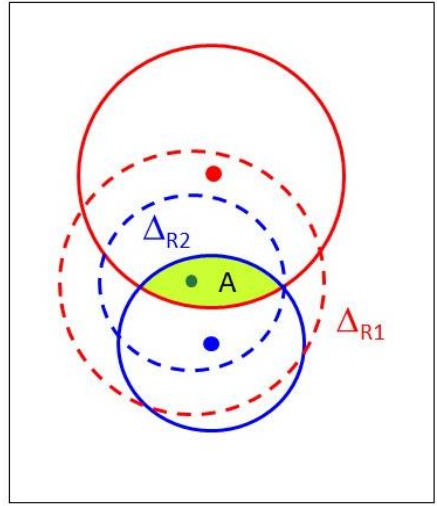

Fig.8. Progressive nucleation: Equivalence between the probability of overlap and the probability of finding nuclei in domains equal to nucleus sizes. The probability a generic point of the space (point $A$ in the figure) belongs to the overlapping region of two nuclei of radius $R_1$ and $R_2$, is equal to the probability that the centers of these nuclei fall within the $\Delta_{R1}$ and $\Delta_{R2}$ domains, namely the dashed circles of radius $R_1$ and $R_2$, centered at $A$.

Let us now consider a collection of D-spheres distributed in size, say $R_1, \dots, R_M$. In this case $p_n$ ($n \neq 0$) *is the probability that a generic point (say A) of the space belongs to the region of overlap of exactly n nuclei. Also in this case $p_n$ is equivalent to the probability that the center of each nucleus of radius $R_i$ lies within the region $\Delta_{R_i}$ of radius $R_i$ centered at A.* Fig.8 displays, graphically, the equivalence between the two definitions. The red and blue points are the centers of red and blue nuclei, the green region is their overlap. The probability that the blue point falls within the blue dashed circle and the red point falls within the red dashed circle, is equivalent to the probability that a generic point of the space (for example point A in the figure) belongs to the green region.

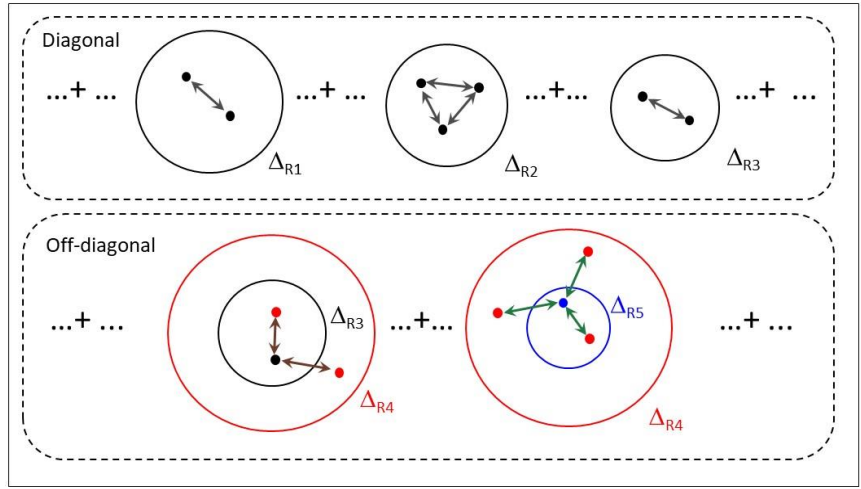

Fig.9. Progressive nucleation: Graphical schematization of eqn.8. In the upper panel, some of the diagonal terms of the series are shown, from the left: $f_{1,1}^{(2)}, f_{2,2,2}^{(3)}, f_{3,3}^{(2)}$. The lower panel refers to some of the off-diagonal terms, from the left: $f_{3,4,4}^{(3)}, f_{5,4,4,4}^{(4)}$. Total number of dots is reported as superscript of the $f$-function. The subscripts are the class indexes: nuclei with the same radius belong to the same class. Please, note the connection between the domain color (circle $\Delta_R$) and the dot color.

In the case of a collection of D-spheres distributed in size, eqn.7 has to be modified by introducing classes (sub-sets) of nuclei with the same size. By denoting with $\alpha_1, \alpha_2, \dots \alpha_m$ the classes, to which corresponds the radii $R_{\alpha_1}, R_{\alpha_2}, \dots, R_{\alpha_m}$ and densities $N_{\alpha_1}, N_{\alpha_2}, \dots, N_{\alpha_m}$, being $\mu$ the total number of classes, it is possible to show that [see Appendix for details]

$$V_{m,ex} = \frac{1}{m!} \sum_{\alpha_1, \alpha_2, \dots \alpha_m}^{\mu} N_{\alpha_1} N_{\alpha_2} \dots N_{\alpha_m} \int_{\Delta_{R_{\alpha_1}}} d\boldsymbol{r}_1$$

$$\times \int_{\Delta_{R_{\alpha_2}}} d\boldsymbol{r}_2 \dots \int_{\Delta_{R_{\alpha_m}}} d\boldsymbol{r}_m f_{\alpha_1, \alpha_2, \dots \alpha_m}^{(m)}(\boldsymbol{r}_1, \dots \boldsymbol{r}_m) , \quad (8)$$

where the sum runs over all classes of dots: $\alpha_i = 1, 2, \dots, \mu$ with $i = 1, 2, \dots, m$ the nucleus label. To explain the meaning of eqn.8 we specialize the equation for $m = 2$. We denote with $f_{\alpha,\beta}^{(2)}$ the 2-nuclei $f$-function for classes $\alpha$ and $\beta$ where $\alpha$ and $\beta$ run from 1 to $\mu$. We get,

$$V_{2,ex} = \frac{1}{2} \sum_{\alpha=1}^{\mu} N_\alpha^2 \int_{\Delta_{R_\alpha}} d\boldsymbol{r}_1 f_{\alpha,\alpha}^{(2)}(\boldsymbol{r}_1, \boldsymbol{r}_2) + \sum_{\alpha>\beta}^{\mu} N_\alpha N_\beta \int_{\Delta_{R_\alpha}} d\boldsymbol{r}_1 \int_{\Delta_{R_\beta}} d\boldsymbol{r}_2 f_{\alpha,\beta}^{(2)}(\boldsymbol{r}_1, \boldsymbol{r}_2), \quad (9a)$$

that is

$$V_{2,ex} = \frac{1}{2} \sum_{\alpha,\beta} N_\alpha N_\beta \int_{\Delta_{R_\alpha}} d\boldsymbol{r}_1 \int_{\Delta_{R_\beta}} d\boldsymbol{r}_2 f_{\alpha,\beta}^{(2)}(\boldsymbol{r}_1, \boldsymbol{r}_2) . \qquad (9b)$$

In Fig.9 some possible configurations for the same classes (diagonal terms) are shown; in particular for two nuclei, three nuclei and two nuclei, of radius $R_1$, $R_2$ and $R_3$, respectively. In the lower panel of Fig.9 some possible configurations for two different classes (off-diagonal terms) are shown, in particular for a nucleus of radius $R_3$ and two nuclei of radius $R_4$; three nuclei of radius $R_4$ and one nucleus of radius $R_5$.

Phase transformation implies a nucleation process which, in turn, gives rise to a size distribution of nuclei, because of the different birth times of nuclei. Eqn.8 can be used to model a time dependent nucleation process by performing a continuum limit. This is done by labeling the classes of nuclei with their birth time, $t_j$, and determining the nucleus radius at running time $t > t_j$. Accordingly, $N_{\alpha_j} \to I(t_j) dt_j$, where $I(t)$ is the nucleation rate. Eqn.8 becomes

$$V_{m,ex}(t) = \frac{1}{m!} \int_0^t I(t_1) dt_1 \dots \int_0^t I(t_m) dt_m \int_{\Delta_{R(t,t_1)}} d\boldsymbol{r}_1$$

$$\times \int_{\Delta_{R(t,t_2)}} d\boldsymbol{r}_2 \dots \int_{\Delta_{R(t,t_m)}} d\boldsymbol{r}_m f_m(\boldsymbol{r}_1, t_1, \dots, \boldsymbol{r}_m, t_m), \qquad (10)$$

where $R(t, t_j)$ is the radius of the sphere, born at time $t_j$, at running time $t$.

Using eqn.10, the transformed volume fraction, eqn.3 is rewritten in term of a series of $f$-functions. These functions depend on both spatial coordinates and birth time of the nuclei. Also, for simultaneous nucleation $I(t) = N\delta(t)$, with $\delta$ Dirac's delta, eqn.10 reduces to eqn.7.

*3.3 g-series*

The series eqns.3,10 can be recast in different form by using the cluster expansion of $f$-functions in terms of correlation functions or $g$-functions:

$$f_1(1) = g_1(1)$$

$$f_2(1,2) = g_1(1)g_1(2) + g_2(1,2)$$

$$f_3(1,2,3) = g_1(1)g_1(2)g_1(3) + g_1(3)g_2(1,2) + g_1(2)g_2(1,3) + g_1(1)g_2(2,3)$$
$$+ g_3(1,2,3) \tag{11}$$

$$\cdot$$
$$\cdot$$
$$\cdot$$

where the arguments of the functions indicate the coordinates of nuclei: $(\boldsymbol{r}_i, t_i)$. Using eqn.11, the $f$-series can be re-summed providing the following series ($g$-series) for the transformed volume [41,42] (see also Appendix for details)

$$V = 1 - \exp\left[\sum_{m=1}^{\infty} \frac{(-)^m}{m!} \int_0^t I(t_1)dt_1 \ldots \int_0^t I(t_m)dt_m\right.$$

$$\times \left.\int_{\Delta_{R(t,t_1)}} d\boldsymbol{r}_1 \ldots \int_{\Delta_{R(t,t_m)}} d\boldsymbol{r}_m \, g_m(\boldsymbol{r}_1, t_1, \ldots, \boldsymbol{r}_m, t_m)\right]. \tag{12a}$$

We stress that in eqn.12a, the quantity $I(t)$ is the nucleation rate[4] where nuclei are subjected to the spatial correlation taken into account by $f$ and $g$- functions.

If $f_m = 1$ , eqn.11 gives $g_m = \delta_{m,1}$, with $\delta_{i,j}$ being the Kronecker delta and eqn.12a becomes the KJMA formula eqn.2. In other words, $I(t)$ contains phantoms, that is the nucleation is random throughout the entire space. We recall that the series eqn.3 was derived using the actual nucleation, whereas $f_m = 1$ implies to compute $V_{k,ex}$ including phantoms. However, it is possible to show that eqn.3 still holds in this case, provided phantoms do not overgrow the transformed phase.

For simultaneous nucleation, $I(t) = N\delta(t)$, eqn.12a reduces to

$$V = 1 - \exp\left[\sum_{m=1}^{\infty} \frac{(-)^m N^m}{m!} \int_{\Delta_R} d\boldsymbol{r}_1 \dots \int_{\Delta_R} d\boldsymbol{r}_m g_m(\boldsymbol{r}_1, \dots, \boldsymbol{r}_m)\right]. \quad (12b)$$

This last equation has been employed in refs.[43,44] for modeling correlated nuclei with hard core interaction and in ref.[45] using a $g_2(r) = e^{-r/\xi}$ where $\xi$ is the correlation length. For $\xi \rightarrow 0$ the genuine Poisson process is obtained. In these works the analytical results have been corroborated by computer simulations.

## 4-Numerical computations

We are now in the position to model phase transformation kinetics without resorting to the concept of phantom. The purpose of this section is to provide two applications of the theory discussed in section 3: i) KJMA-compliant transformations; ii) transformations where phantom overgrowth takes place.

### 4.1 Modeling KJMA-compliant transformations through series

As anticipated in the introduction, in any phase change by non-simultaneous nucleation and growth, actual nucleation is a correlated process as it can only occur in the untransformed phase. This is also true for a KJMA-compliant transformation with respect to the actual nucleation.

---

[4] $I(t)dt$ is the number of nuclei, per unit volume, which start growing within $dt$ at $t$.

In this sub-section we show that eqn.12a, for KJMA-compliant transformations, leads to the KJMA formula, eqn.2.

The relationship between actual ($I_a$) and phantom-included ($I_p$) nucleation rate reads $I_a(t) = I_p(t)\big(1 - V(t)\big)$, which holds for any transformation in homogeneous systems. For a KJMA-compliant transformation it become

$$I_a(t) = I_p(t)e^{-\int_0^t I_p(t')v(t-t')dt'}, \qquad\qquad (13)$$

to be used in eqn.12a. As far as the correlation is concerned, we made use of the approach proposed by Kirkwood [39] based on superposition of $f_2$ functions. The correlation constraint implies that an actual nucleus can only form in the untransformed volume. This means that two (actual) nuclei with birth times $t_1$ and $t_2$ (with $t_1 > t_2$) have to be located at relative distance greater that $R(t_1-t_2)$. In this case the nucleus that born at time $t_1$ cannot be a phantom. The simplest form for $f_2$ is $f_2(\boldsymbol{r}_1, t_1, \boldsymbol{r}_2, t_2) = H(|\boldsymbol{r}_1 - \boldsymbol{r}_2| - R(t_1-t_2))$, being $H(x)$ the Heaviside step function ($H(x) = 1, x > 0$; $H(x) = 0, x < 0$). The present form for $f_2$ describes a time dependent hard sphere interaction, which is modeled through the potential well similar to that of Fig.1. Higher order $f$-functions are attained from the superposition of $f_2$ functions. For instance, the $f_3$-function becomes: $f_3(1,2,3) = f_2(1,2)f_2(1,3)f_2(3,2)$ and, similarly, $f_m$ as the product of $\binom{m}{2}$ $f_2$ terms. It is worth noticing that the $f_2$ function is given by a series in terms of Mayer's function[5]; the Heaviside function is the lowest order term of the expansion. As far as the $g$-correlation functions are concerned, their computation is done in terms of $f$-functions by means of eqn.11. For instance, $g_2(1,2) = f_2(1,2)-f_1(1)f_1(2)$, next $g_3(1,2,3)$ is computed in terms of $f_1$, $f_2$ and $f_3$ using the last expression of eqn.11 by inserting the $g_2$ expression in terms of $f_1$ and $f_2$. Using $f$-functions we computed both $f$ and $g$- series for the transformed volume in the case of constant nucleation rate $I_p$ in eqn.13 and linear growth of nuclei. The computation has been performed for D=1, 2, 3 for which the extended volume in eqn.2 reads:

---

[5] In statistical thermodynamics Mayer's function is defined as $\tilde{f} = e^{-\beta v(r)} - 1$, where $v(r)$ is the couple interaction potential, $r$ relative distance and $\beta = 1/k_B T$. For $v(r)_{r \leq \sigma} = \infty$, and $v(r)_{r > \sigma} = 0$, $\tilde{f} = H(r - \sigma) - 1$ where $e^{-\beta v(r)} = H(r - \sigma)$ is the lowest order term of $f_2$.

$$V_e(t) = I_p \int_0^t v(t-t')dt' = I_p\omega_D v^D \frac{t^{D+1}}{D+1}, \qquad (14)$$

where $\omega_D$ is the geometrical factor ($\omega_1 = 2$; $\omega_2 = \pi$; $\omega_3 = \frac{4}{3}\pi$) and v the linear growth rate, $R(t) = vt$. Note that $V_e$ coincides with $V_{1,ex}$ of eqn.3 with the inclusion of phantoms. In what follows, we illustrate the computation of $f_2$ -containing terms in the $f$-series, namely $V_{2,ex}$ (eqn.3) which takes a simple form for KJMA-compliant transformations. Similar terms also enter the $g$-series. The homogeneity of the system allows using relative coordinates in eqn.10, at $m = 2$, according to

$$V_{2,ex}(t) = I_p^2 \int_0^t dt_1 e^{-V_e(t_1)} \int_0^{t_1} dt_2\, e^{-V_e(t_2)} \int_{\Delta_{R(t,t_1)}} d\boldsymbol{r}_1 \int_{\Delta_{R(t,t_2)}} d\boldsymbol{r}\, H(r - R(t_1-t_2)), \quad (15)$$

where $r = |\boldsymbol{r}_1 - \boldsymbol{r}_2|$ and the integral has been rewritten with time ordering being the integrand symmetric under time variable exchange. Let us focus our attention on the double integral over space variables. The $\boldsymbol{r}_1$ variable spans the domain $\Delta_{R(t,t_1)} \subseteq \Delta_{R(t,t_2)}$ (Fig.10 panel a); at a given $\boldsymbol{r}_1$, owing to the Heaviside function, the $\boldsymbol{r}$ variable spans the region given by the difference $\Delta_{R(t,t_2)} - \Delta_{R(t_1,t_2)}$ (dashed region in Fig.10a) where $\Delta_{R(t_1,t_2)}$ is the D-sphere centered at $\boldsymbol{r}_1$ with radius $R(t_1,t_2) = R(t_1 - t_2)$. The important point is that for linear growth, "correlation D-sphere" is entirely contained within $\Delta_{R(t,t_2)}$ (see Fig.10 panel a). In general, convex $R(t)$ functions fulfil the following condition: $\Delta_{R(t_1,t_2)} \subseteq \Delta_{R(t,t_2)}$. Computation of the space integral for linear growth is immediate and provides

$$\omega_D v^D \int_{\Delta_{R(t,t_1)}} d\boldsymbol{r}_1\, [(t-t_2)^D - (t_1-t_2)^D] = (\omega_D v^D)^2 (t-t_1)^D[(t-t_2)^D - (t_1-t_2)^D]. (16)$$

On the other hand, $\Delta_{R(t_1,t_2)} \not\subset \Delta_{R(t,t_2)}$ holds for concave $R(t)$ functions. These functions cannot be used in KJMA approach since they give rise to phantom overgrowth. In this case the overlap between the two domains have to be duly considered for estimating $V_{2,ex}$ [46,47] (see Fig.10 panel b).

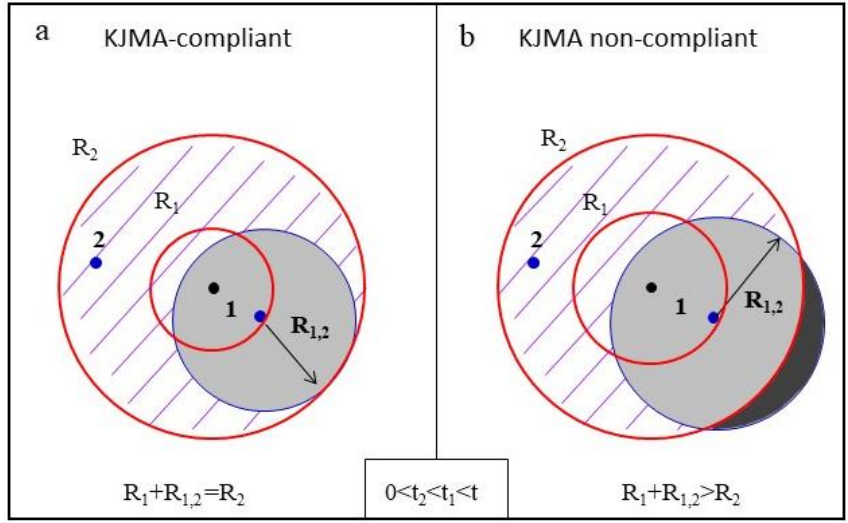

Fig.10. The dashed region is the integration domain for the nucleus 2 (birth time $t_2$) in the case of hard-core pair interaction between nuclei 1 and 2. The grey region is prohibited to nucleus 2 and for KJMA compliant transformations is entirely inside the disk of radius $R_2 = R(t - t_2)$ (panel a); whereas in the case of concave growth law it protrudes out of the $R_2$ disk. In the figure $R_{1,2} = R(t_1 - t_2)$ is the correlation disk.

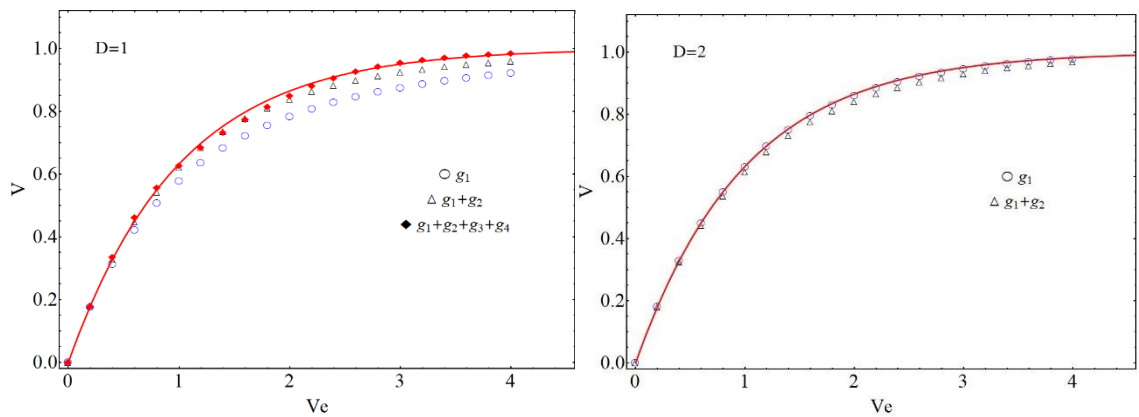

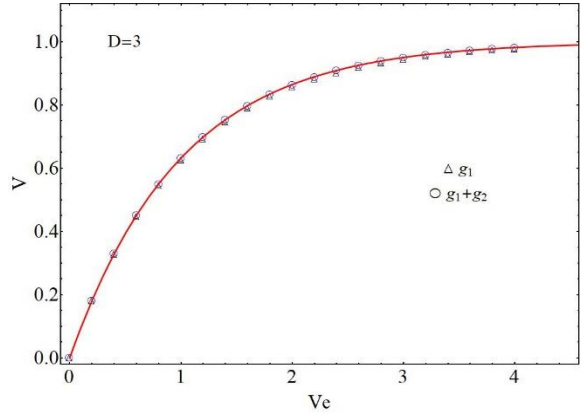

Fig.11. Kinetics of KJMA-compliant transformation in terms of actual nuclei. To this purpose eqn.12a has been computed using actual nucleation rate and hard-core interaction potential. The output of the computation for D=1,2,3 and for various truncation of the g-series are shown as open symbols. Solid line is the exact kinetics eqn.2, for linear growth and constant nucleation rate, $I_p$.

In Fig.11 we report the outputs of the computation of eqn.12a for D=1-3 for various truncations of the $g$-series (open symbols). In the same plots the solid line is the KJMA formula, eqn.2, that is the exact kinetics. The kinetics are displayed as a function of $V_e$ that is the argument of the exponential in eqn.2. The computation for 2D system has been proposed in previous work [48] while those for 1D and 3D systems are original results. To obtain a satisfactory agreement with the KJMA formula, while in 2D and 3D only two terms are enough in the expansion of $g$, for 1D it takes four. The greater the number of phantoms, the greater the number of terms needed to describe the kinetics. It is quite intuitive that, with the same density of nuclei, phantom nuclei are more likely to form in 1D than in 2D or 3D; in fact, in 1D there is less space available for nucleation in untransformed space. In fig.12 the behavior of the number of phantom nuclei as a function of the extended volume is displayed.

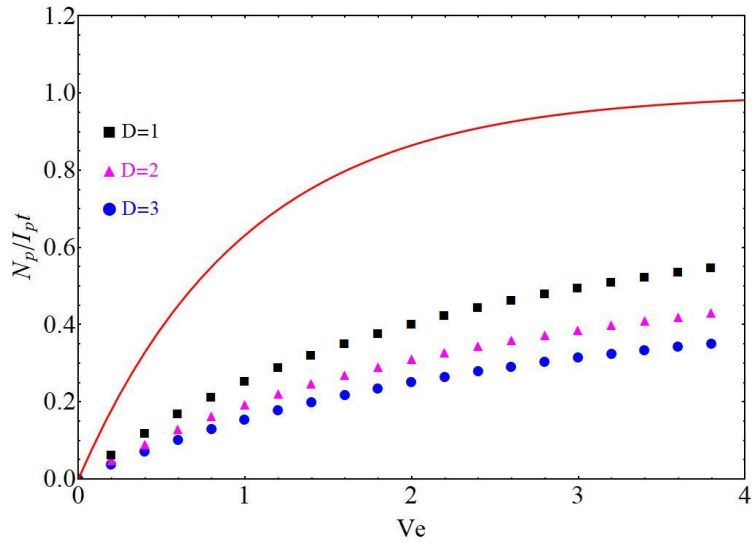

Fig.12. Behavior of the number of phantom nuclei, as a function of extended volume, in the case of linear growth and constant nucleation rate for D=1,2,3. The number density of phantom nuclei is normalized to the total density of nucleation, namely $I_p t$.

As far as the $f$-series is concerned, in Fig.13 we report the various contributions for the 1D system. It is evident that, at the same order of approximation, the $g$-series works much better. This is not surprising since $g$-series was obtained by partially summing the $f$-series.

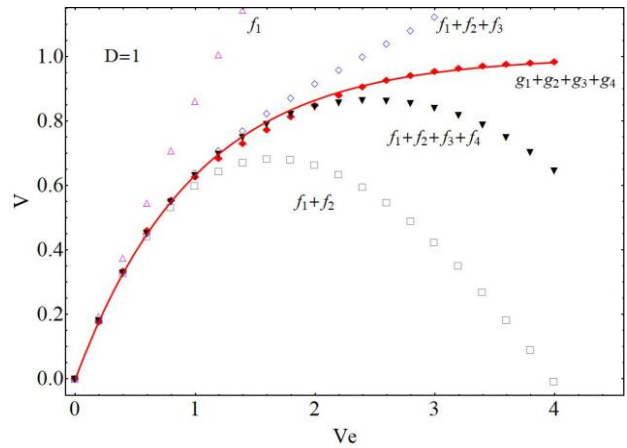

Fig.13. 1D-kinetics of KJMA compliant transformations (eqn.3,10) through $f$-series. The various orders of approximation are displayed as symbols. The kinetics obtained by truncation of the $g$-series up to $g_4$, and $f$-series up to $f_4$ are also reported as full symbols. Solid line is the KJMA solution.

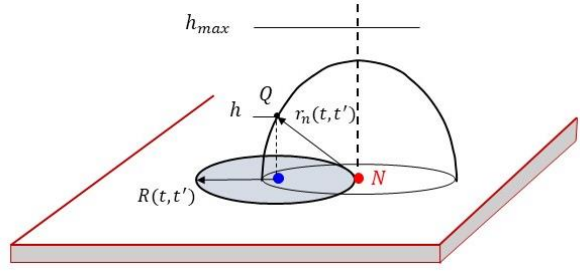

Fig.14 Schematic representation of a hemispherical nucleus on a solid surface, with growth law $r_n(t, t')$. The generic point of the space, $Q$, located at height $h$, is not transformed at time $t$ provided no nucleation event occurs between $t'$ and $t' + dt'$ within the colored disk (radius $R^2(t, t') = r_n^2 - h^2$). In fact, nucleation events occurring within the disk in time interval $t' - t' + dt'$ are capable of transforming $Q$ before time $t$. The stochastic problem is therefore equivalent to a stochastic process of dots in 2D-space. The hemisphere centered at $N$ represents a nucleus just at the border of the disk. $h_{max} \equiv r_n(t, 0)$ is the maximum height of the nucleus at $t$.

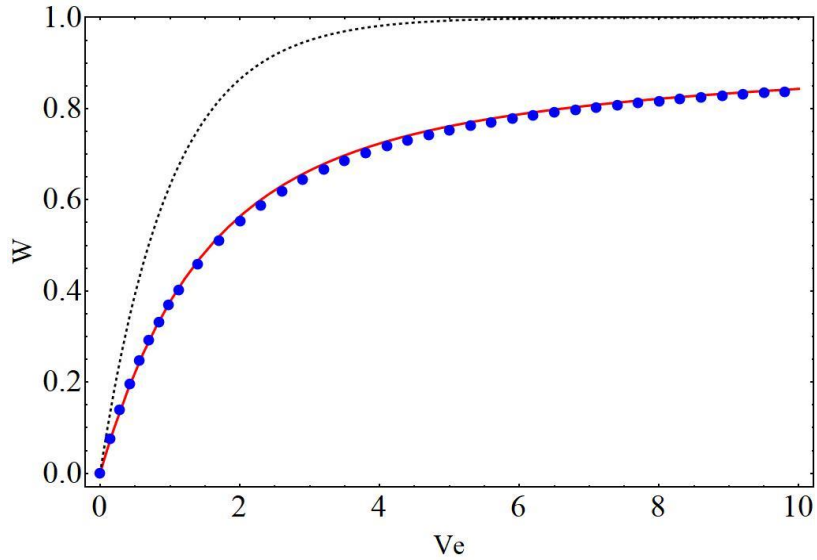

Fig.15 Kinetics of film deposition at solid surface in the case of hemispherical nuclei with diffusional growth and constant nucleation rate ($I_p$). In the plot, $W$ is the volume of the film, per unit area, normalized to the maximum radius of the nucleus, $h_{max} \equiv r_n(t, 0)$. $V_e$ is the phantom included extended volume for constant nucleation rate. The solution attained in ref.[12] by using the KJMA model and the kinetics computed using the $g$-series in terms of actual nucleation rate [49] are shown as solid line and full symbols, respectively. The behavior of the fraction of substrate surface covered by the film is displayed as dotted line. It follows that the kinetics is well representative of the closure of the film. The computation of the actual nucleation rate was approximated by eqn.13.

Phase transformations in terms of actual nucleation rate has been modeled in the case of 3D-nucleus growth on solid surfaces. In this context, an important example is the electrodeposition of a new phase at electrode surface [49]. It was originally called 2D-1/2 growth mode, to stress that just 1/2 of the third dimension (positive z axis) is involved in the film deposition (Fig.14). The general theory of electrodeposition by nucleation and parabolic growth law, has been proposed by Bosco et al in ref.[12] on the basis of the KJMA model for phase change in 2D space. In ref.[49] the approach based on correlation functions was employed to describe the same system by employing actual nucleation rate and the g-series. Besides, it allows to estimate the contribution of phantom overgrowth. The result is displayed in Fig.15 (solid symbols) together with the KJMA kinetics which implies phantom nucleation (see Appendix for details). A good agreement between the two curves is attained by computing the g-series up to the second order terms [49]. In fact, the effect of phantom overgrowth is found to be small (about 2%) and this justifies the application of the KJMA theory in reference [12].

### 4.2 Phantom overgrowth

As discussed in section 3.1 Avrami's series also holds by allowing nucleation to occur even in the transformed space, provided phantoms do not protrude out of the real nuclei. We recall that inclusion of phantoms is needed in KJMA theory to deal with a genuine Poisson process. Therefore, phantom overgrowth is strictly linked to growth laws [35]. In particular, it is worth repeating what was said in section 2: concave functions give rise to the phenomenon of overgrowth. The paradigmatic case is the diffusional growth, where $R(t) \propto \sqrt{t}$ or $\frac{dR}{dt} \propto \frac{1}{R}$. An example of concave growth laws is shown in Fig.16 for two power functions, i.e. $r(t) = \sqrt{t}$ and $r(t) = \sqrt[4]{t}$. In the figures, the black and red points are an actual and a phantom dot, respectively, the growth kinetics are represented with the respective colors. The overgrowth starts at the intersection point between the red and black curves. Note that the overgrowth becomes the more important the lower the power exponent. The KJMA compliant $R(t)$ functions must be convex.

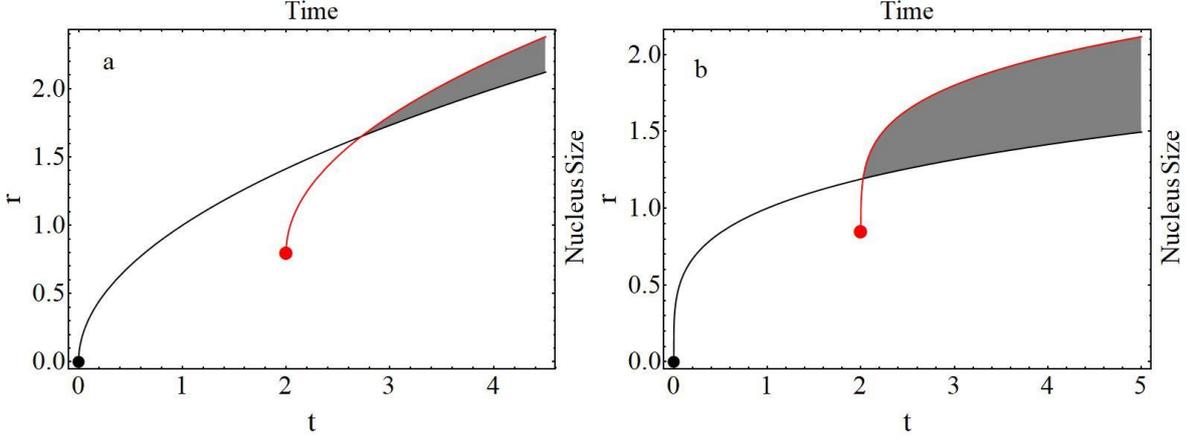

Fig.16. Pictorial view of phantom overgrowth for concave growth laws: panel a) $R(t) = \sqrt{t - t'}$, panel b) $R(t) = \sqrt[4]{t - t'}$. Each plot displays the behavior of the nucleus boundary for two nuclei, $r(t)$, referred to the center of the black nucleus. Specifically, for the black nucleus $r(t) = \sqrt{t}$ and $r(t) = \sqrt[4]{t}$, whereas for the red one $r(t) = \sqrt{t-2} + 0.8$ and $r(t) = \sqrt[4]{t-2} + 0.8$. The shaded region indicates the evolution of the overgrowth in time-space frame.

However, phantom overgrowth can be healed by dealing with the actual nucleation at the cost of considering spatial correlation. To clarify the point, let us consider again the second order term of Avrami's series, $V_{2,ex}(t)$, according to eqn.15. The integrand over the relative coordinate of the second nucleus, $\boldsymbol{r}$, requires, for some configurations, considering the volume of the correlation sphere (of radius $R(t_1 - t_2)$) protruding out of the sphere of radius $R(t - t_2)$, as displayed in panel b of Fig.10. In fact, for a concave $R(t)$ the inequality $R(t - t_1) + R(t_1 - t_2) > R(t - t_2)$, with $0 < t_2 < t_1 < t$, is always satisfied. In addition, the actual nucleation rate is not given, in this case, by the KJMA equation (eqn.2). In the simplest case of constant $I_p$, $I_a(t) = \left(1 - V(t)\right)I_p$ which leads, once inserted in $f$ and $g$- series, to a very complex integral equation for $V(t)$. However, the effect of overgrowth in diffusion-type growth has been shown to be negligible [49-52]; as a first approximation $I_a(t)$ can be estimated using the KJMA equation [49].

An interesting approach for dealing with phantom overgrowth in case of Poissonian nucleation (in the untransformed phase) has been proposed by Alekseechkin [52] based on the so-called "critical region method". In this method, the rate of phase transformation is expressed according to

$$\frac{dV}{dt} = D\omega_D [1 - V(t)] \int_0^t q(t_1|t) \, I_p\,(t) R^{D-1}(t_1,t) \partial_t R(t_1,t) dt_1]. \quad (17)$$

In eqn.17, the term $[1 - V(t)]$ is the probability a generic point (say point c) of the system is not transformed until time $t$, $I_p\,(t)$ is the phantom included nucleation rate, $q(t_1|t)$ is the conditional probability that the nucleation point (at relative distance $R(t_1,t)$ from the point c) is not transformed until $t_1$ provided c is untransformed until $t$. In this way phantom overgrowth does not contribute to the transformation. For KJMA compliant transformation $q(t_1|t) = 1$ and eqn.17 reduces to eqn.2.

Using geometrical argument, the author derived the following equation for the $q$ probability [52]:

$$q(\tau|t) = \exp\left[ -\int_0^\tau dt_1 \int_0^{t_1} dt_2 \, q(\tau'|\tau) \, I(t_2) \partial_{t_1} \Omega_D(t_1,t_2,t) dt_2 \right] \quad (18)$$

where $\Omega_D(t_1,t_2,t)$ is the portion of the volume of the sphere of radius $R(t_1 - t_2)$ protruded from that of radius $R(t - t_2)$, the distance between these two spheres is $R(t - t_1)$ (see also Fig.10b). By solving eqn.18 by successive approximations, the effect of phantom overgrowth on the kinetics has been estimated in the case of parabolic growth [52]. Eqn.18 can be traced back to the method based on correlation functions, namely the $g$-series discussed in section 3.2.2 (eqn.12a). By comparing the time derivative of eqn.12a with eqn.17 it is possible to determine the functional form of $q(t_1|t)$ in terms of correlation functions. This issue has been discussed in a certain detail in ref.[46]. Using analytical methods and computer simulations, the effect of phantom overgrowth, as the difference between real and KJMA kinetics, has been estimated to be lower than 0.02 for both 2D and 3D transformations [46,47].

A transformation mechanism where phantom overgrowth is quite significant, is that ruled by non-simultaneous nucleation with instantaneous growth of nuclei to their final size $R$. This mechanism, which is not compliant with the KJMA model, has been shown to be relevant in crystal growth taking place in thin layer between two interfaces [53,54]. In the 2D case, an equivalent process to this mechanism would be to throw disks at random onto a flat surface removing any disk whose center falls into an area occupied by previously thrown disk. In other words, removing all phantoms. This problem has been tackled by Tobin [55] and later by Tagami et al [54] who revised the KJMA model to account for overgrowth. In ref.[54] the authors proposed a rate equation for the transformed volume fraction by modeling the rate of overgrowth introducing the so called overlap parameter, $\gamma$. It is defined as the average value of the overlap volume (normalized to nucleus volume) between a phantom and an actual nucleus. It follows that the total volume of the region protruding the transformed phase equals $(1 - \gamma)[1 - V(t)]$. On the other hand, the changing rate of the total volume of phantoms is $I_p v \, V(t)$ where $v$ is the volume of the D-sphere of radius $R$ and $I_p$ is the constant nucleation rate (phantom included). Therefore, the total contribution to the overgrowth reads $I_p v \, V(t)(1 - \gamma)[1 - V(t)]$. In the present case the growth is instantaneous and the KJMA equation (eqn.2) would imply $V(t) = 1 - e^{-\int_0^t I_p v_D dt'} = 1 - e^{-I_p v t}$ that is the rate equation $\frac{dV}{dt} = I_p v \left(1 - V(t)\right)$. The kinetics is eventually computed by the authors by subtracting the contribution of phantom overgrowth from the KJMA rate equation:

$$\frac{dV}{dt} = I_p v \ (1 - V) - I_p v \ V(1 - \gamma)[1 - V] \ . \quad (19a)$$

In terms of $V_e(t) = I_p v t$, eqn.19a can be recast as,

$$\frac{dV}{dV_e} = \ [1 - V][1 - V(1 - \gamma)] \quad (19b)$$

whose solution is

$$V = \frac{e^{\gamma V_e} - 1}{e^{\gamma V_e} - 1 + \gamma} \ . \quad (20)$$

Another approach developed in the middle of the 1950s with the purpose of finding a more general kinetic equation, is based on the introduction of a parameter, named "impingement factor", $\eta$, in the differential form of the KJMA equation [56-58]. In formula

$$\frac{dV}{dV_e} = (1 - V)^\eta, \qquad (21)$$

with solution given by ($\eta \neq 1$),

$$V(V_e) = 1 - \frac{1}{[1 + V_e(\eta - 1)]^{1/(\eta-1)}} . \qquad (22)$$

For $\eta = 1$ eqn.21 reduces to the KJMA equation, whereas for $\eta = 2$ it reduces to the Austin Rickett equation [59]. In order to gain insight into the meaning of the impingement factor, in ref.[60] eqn.21 has been studied in the framework of the correlation function approach. The analysis is based on the comparison between Taylor's expansion of eqn.22 in terms of $V_e$, and the $f$-series eqns.3,10. In the $f$-series use was made of the actual nucleation rate with a hard disk correlation potential; in particular, at the lowest order $f_2(\boldsymbol{r_1}, \boldsymbol{r_2}) = H(|\boldsymbol{r_1} - \boldsymbol{r_2}| - R)$. The detailed analysis of eqn.22 leads to the following expression for $\eta$: $\eta = 2 - \gamma$. Notably, the important output of the analysis is the relationship between the overlap parameter and the pair-correlation function. In the 2D case the computation provides $\gamma = 1 - \frac{3^{\frac{3}{2}}}{4\pi} \cong 0.86$ [54,60].

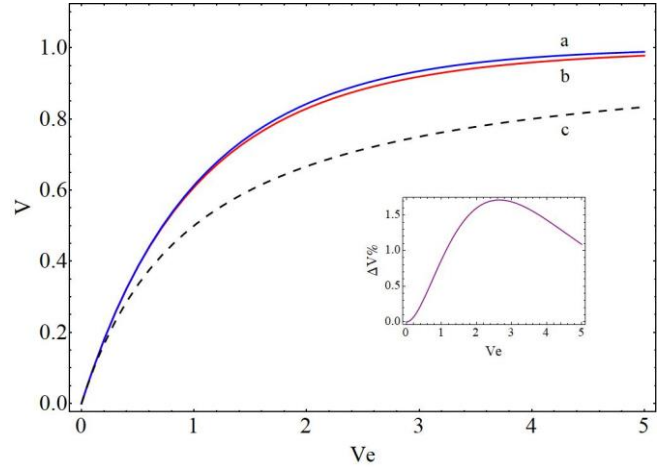

Fig.17. Kinetics of transformation by progressive nucleation and instantaneous growth (2D case). Curve a) is the output of the model developed by Tagami et al [54] by means of eqn.20; curve b) is the kinetics obtained using the differential equation eqn.21 with $\eta = 2 - \gamma$ and $\gamma = 1 - \frac{3^{\frac{3}{2}}}{4\pi}$. Curve c) is the kinetics for $\eta = 2$ [55]. Inset: percentage difference between curve a) and b).

The two approaches which give eqns.20, 22, in fact provides the same kinetics, as shown in Fig.17 for 2D transformation, being the difference lower than 1.5%.

## 5- Conclusions

In this minireview, we have made an excursus about the importance of the concept of phantom in the KJMA model and how to overcome the constraints that this concept entails. In particular it has been underlined that the introduction of an unphysical quantity such as phantom, makes the determination of the kinetics much simpler but, if associated with a concave growth law, it gives rise to the phenomenon of overgrowth that invalidates the kinetic function.

Sinoptically:

1. We emphasized that Avrami's series to get transformed volume (eqn.3 of this paper) describes correlated nucleation. It has been derived through a geometrical analysis (set theory) of the overlaps among nuclei. Furthermore, the connection between the geometry of the overlaps and the statistical mechanics of dots (particles) has been established in terms of the *f*-functions (Van Kampen's notation).

2. To get the KJMA formula without resorting to the subtle concept of phantom, it is mandatory to use correlation functions. In particular, we have shown that the pair correlation function (Van Kampen's notation) is enough for D=2, 3, whereas expansion up to $g_4$ is needed for the 1D case.

3. Using diffusional growth law $\left(R \sim t^{\frac{1}{2}}\right)$ into KJMA model, the effect of overgrowth gives a negligible contribution to the real kinetics. On the other hand, in the Tobin model, where overgrowth is important, KJMA formula does not work. In this case, a modified differential form of the KJMA equation has been proposed by considering an empirical parameter: the *impingement factor*. However, using the *f*-series expansion it was shown that this quantity is directly linked to the pair-distribution function for hard core interaction.

**Acknowledgement**

The Authors are grateful to Giulio Fanfoni for having realized figures 3, 4 and 5.

**Appendix**

*1-Derivation of eqn.8*

The $V_{3,ex}$ term is given by the sum of the following three contributions,

$$V_{3,ex} = \frac{1}{2!} \sum_{\alpha \neq \beta} N_\alpha^2 N_\beta \int_{\Delta_{R_\alpha}} d\boldsymbol{r}_1 \int_{\Delta_{R_\alpha}} d\boldsymbol{r}_2 \int_{\Delta_{R_\beta}} d\boldsymbol{r}_3 f_{\alpha,\alpha,\beta}^{(3)}$$

$$+ \sum_{\alpha > \beta > \gamma} N_\alpha N_\beta N_\gamma \int_{\Delta_{R_\alpha}} d\boldsymbol{r}_1 \int_{\Delta_{R_\beta}} d\boldsymbol{r}_2 \int_{\Delta_{R_\gamma}} d\boldsymbol{r}_3 f_{\alpha,\beta,\gamma}^{(3)} +$$

$$\frac{1}{3!} \sum_\alpha N_\alpha^3 \int_{\Delta_{R_\alpha}} d\boldsymbol{r}_1 \int_{\Delta_{R_\alpha}} d\boldsymbol{r}_2 \int_{\Delta_{R_\alpha}} d\boldsymbol{r}_3 f_{\alpha,\alpha,\alpha}^{(3)}, \qquad (A1)$$

where the factorial terms 2! and 3! correct for the equivalent configurations of $\alpha\alpha$ couples and $\alpha\alpha\alpha$ triplets, respectively. Eqn. A1 can be rewritten as

$$V_{3,ex} = \frac{1}{2!} \sum_{\alpha \neq \beta} N_\alpha^2 N_\beta \frac{1}{3} \left[ \int_{\Delta_{R_\alpha}} d\boldsymbol{r}_1 \int_{\Delta_{R_\alpha}} d\boldsymbol{r}_2 \int_{\Delta_{R_\beta}} d\boldsymbol{r}_3 \left( f_{\alpha,\alpha,\beta}^{(3)} + f_{\alpha,\beta,\alpha}^{(3)} + f_{\beta,\alpha,\alpha}^{(3)} \right) \right]$$

$$+ \frac{1}{3!} \sum_{\alpha \neq \beta \neq \gamma \neq \alpha} N_\alpha N_\beta N_\gamma \int_{\Delta_{R_\alpha}} d\boldsymbol{r}_1 \int_{\Delta_{R_\beta}} d\boldsymbol{r}_2 \int_{\Delta_{R_\gamma}} d\boldsymbol{r}_3 f_{\alpha,\beta,\gamma}^{(3)}$$

$$+ \frac{1}{3!} \sum_\alpha N_\alpha^3 \int_{\Delta_{R_\alpha}} d\boldsymbol{r}_1 \int_{\Delta_{R_\alpha}} d\boldsymbol{r}_2 \int_{\Delta_{R_\alpha}} d\boldsymbol{r}_3 f_{\alpha,\alpha,\alpha}^{(3)},$$

that is

$$V_{3,ex} = \frac{1}{3!} \sum_{\alpha,\beta,\gamma} N_\alpha N_\beta N_\gamma \int_{\Delta_{R_\alpha}} d\boldsymbol{r}_1 \int_{\Delta_{R_\beta}} d\boldsymbol{r}_2 \int_{\Delta_{R_\gamma}} d\boldsymbol{r}_3 \int f_{\alpha,\beta,\gamma}^{(3)} \qquad (A2)$$

where the sum runs over all the $\mu$-classes. The general term of the extended volume of order $k$ therefore reads

$$V_{k,ex} = \frac{1}{k!} \sum_{\alpha_1,\alpha_2,\ldots,\alpha_k} N_{\alpha_1} N_{\alpha_2} \ldots N_{\alpha_k} \int_{\Delta_{R_{\alpha_1}}} d\boldsymbol{r}_1 \int_{\Delta_{R_{\alpha_2}}} d\boldsymbol{r}_2 \ldots \int_{\Delta_{R_{\alpha_k}}} d\boldsymbol{r}_k \, f^{(k)}_{\alpha_1,\alpha_2,\ldots,\alpha_k} \quad , \quad (A3)$$

where $\alpha_i = 1,2,\ldots,\mu$ is the class index with $i = 1,2,\ldots,k$ the nucleus label.

## 2- Derivation of the g-series

The cluster expansion eqn.11 can be written according to

$$f_m(\boldsymbol{r}_1,\ldots,\boldsymbol{r}_m) = \sum_{\boldsymbol{n}} \sum_{P} [g_1]^{n_1} [g_2]^{n_2} \ldots [g_m]^{n_m} \quad , \quad\quad (A4)$$

where the components of $\boldsymbol{n}$ satisfies $\sum_k k n_k = m$ and $P$ indicates that only distinct contribution arising from the permutations of the $\boldsymbol{r}_1,\ldots,\boldsymbol{r}_m$ variables, in the product of the $g$'s, have to be retained. In eqn.A4 we made use of a short notation for the terms of the sum: for instance, $[g_1]^2 = g_1(1)g_1(3)$, $[g_2]^3 = g_2(4,7)g_2(6,5)g_2(2,8)$, and so on. By means of eqn.A4 eqn.7 provides

$$V_{m,ex} = \frac{N^m}{m!} \sum_{\boldsymbol{n}} \sum_{P} \int_{\Delta_R} [g_1]^{n_1}[g_2]^{n_2} \ldots [g_m]^{n_m} d\boldsymbol{r}_1 \ldots d\boldsymbol{r}_m =$$

$$\frac{N^m}{m!} \sum_{\boldsymbol{n}} \sum_{P} \left( \int_{\Delta_R} g_1 d\boldsymbol{r}_1 \right)^{n_1} \left( \int_{\Delta_R} d\boldsymbol{r}_1 \int_{\Delta_R} d\boldsymbol{r}_2 \, g_2 \right)^{n_2} \ldots \left( \int_{\Delta_R} d\boldsymbol{r}_{3\ldots} \int_{\Delta_R} d\boldsymbol{r}_m g_m \right)^{n_m} . \ (A5)$$

For a given sequence $\boldsymbol{n} = (n_1, n_2, \ldots, n_m)$, the number of equivalent terms entering the $P$-sum of eqn.A5 is equal to $\frac{m!}{[1!]^{n_1}[2!]^{n_2}\ldots[m!]^{n_m}\, n_1! n_2! \ldots n_m!}$. Eqn.A5 provides

$$V_{m,ex} = \frac{N^m}{m!} \sum_{\boldsymbol{n}} m! \frac{\left(\frac{1}{1!}\int_{\Delta_R} g_1 d\boldsymbol{r}_1\right)^{n_1}}{n_1!} \frac{\left(\frac{1}{2!}\int_{\Delta_R} d\boldsymbol{r}_1 \int_{\Delta_R} d\boldsymbol{r}_2 \, g_2\right)^{n_2}}{n_2!} \ldots \frac{\left(\frac{1}{m!}\int_{\Delta_R} d\boldsymbol{r}_{3\ldots} \int_{\Delta_R} d\boldsymbol{r}_m g_m\right)^{n_m}}{n_m!}$$

and the series eqn.3 becomes

$$V = \sum_{m=1}^{\infty} (-)^{m+1} \frac{N^m}{m!}$$

$$\times \sum_{\boldsymbol{n}} m! \frac{\left(\frac{1}{1!}\int_{\Delta_R} g_1 d\boldsymbol{r}_1\right)^{n_1}}{n_1!} \frac{\left(\frac{1}{2!}\int_{\Delta_R} d\boldsymbol{r}_1 \int_{\Delta_R} d\boldsymbol{r}_2\, g_2\right)^{n_2}}{n_2!} \dots \frac{\left(\frac{1}{m!}\int_{\Delta_R} d\boldsymbol{r}_{3..} \int_{\Delta_R} d\boldsymbol{r}_m g_m\right)^{n_m}}{n_m!} \,,$$

that is rewritten as

$$V = 1 - \sum_{m=0}^{\infty} (-N)^{(1\, n_1 + 2\, n_2 + \cdots m\, n_m)}$$

$$\times \sum_{n_1, n_2, \dots, n_m} \frac{\left(\frac{(-N)}{1!}\int_{\Delta_R} g_1 d\boldsymbol{r}_1\right)^{n_1}}{n_1!} \frac{\left(\frac{(-N)^2}{2!}\int_{\Delta_R} d\boldsymbol{r}_1 \int_{\Delta_R} d\boldsymbol{r}_2\, g_2\right)^{n_2}}{n_2!} \dots \frac{\left(\frac{(-N)^m}{m!}\int_{\Delta_R} d\boldsymbol{r}_1 \dots \int_{\Delta_R} d\boldsymbol{r}_m g_m\right)^{n_m}}{n_m!}. \quad (A6)$$

Because of the sum over $m$, the $n_1, n_2, \dots, n_m$ values in the second sum run, from 0 to $\infty$ independently. Eqn.A6 gives eventually

$$V = 1 - \exp\left[\sum_{m=1}^{\infty} (-)^m \frac{N^m}{m!} \int_{\Delta_R} d\boldsymbol{r}_1 \dots \int_{\Delta_R} d\boldsymbol{r}_m g_m(\boldsymbol{r}_1, \dots, \boldsymbol{r}_m)\right] \quad (A7)$$

that is eqn.12b.

A similar expression can be developed in the case of size distributed D-sphere. We start from eqn.10 and made use of the cluster expression above (eqn.A4) where the numbers in the argument of the $g$ functions now label nucleus coordinates $(\boldsymbol{r}_i, t_i)$. It follows

$$V_{m,ex}(t) = \frac{1}{m!} \sum_{\boldsymbol{n}} \sum_{P} \int_0^t I(t_1) dt_1 \dots \int_0^t I(t_m) dt_m \int_{\Delta_{R(t,t_1)}} d\boldsymbol{r}_1$$

$$\times \int_{\Delta_{R(t,t_2)}} d\boldsymbol{r}_2 \dots \int_{\Delta_{R(t,t_m)}} d\boldsymbol{r}_m \, [g_1]^{n_1}[g_2]^{n_2} \dots [g_m]^{n_m}. \quad (A8)$$

Following the same computation pathway as above, one obtains,

$$V_{m,ex}(t)$$

$$= \sum_{\boldsymbol{n}} \frac{\left(\frac{1}{1!}\int_0^t I(t_1)dt_1 \int_{\Delta_{R(t,t_1)}} g_1 d\boldsymbol{r}_1\right)^{n_1}}{n_1!} \; \frac{\left(\frac{1}{2!}\int_0^t I(t_1)dt_1 \int_0^t I(t_2)dt_2 \int_{\Delta_{R(t,t_1)}} d\boldsymbol{r}_1 \int_{\Delta_{R(t,t_2)}} d\boldsymbol{r}_2 \, g_2\right)^{n_2}}{n_2!} \dots$$

$$\frac{\left(\frac{1}{m!}\int_0^t I(t_1)dt_1 \dots \int_0^t I(t_m)dt_m \int_{\Delta_{R(t,t_1)}} d\boldsymbol{r}_1 \dots \int_{\Delta_{R(t,t_m)}} d\boldsymbol{r}_m g_m\right)^{n_m}}{n_m!}$$

that leads, once inserted in eqn.3, to the $g$-series expression of the transformed volume:

$$V(t)$$

$$= 1 - \exp\left[\sum_{m=1}^{\infty} (-)^m \frac{1}{m!} \int_0^t I(t_1)dt_1 \dots \int_0^t I(t_m)dt_m \int_{\Delta_{R(t,t_1)}} d\boldsymbol{r}_1 \dots \int_{\Delta_{R(t,t_m)}} d\boldsymbol{r}_m \, g_m\right], \quad (A9)$$

where $g_m \equiv g_m(\boldsymbol{r}_1, t_1, \dots, \boldsymbol{r}_m, t_m)$.

### 3- KJMA approach to 2D-1/2 growth mode

To compute the volume of the deposit we determine the probability, $P(t,h)$, that a generic point at height $h$ from the substrate, is not transformed by the new phase within time $t$ (Fig.15). This implies that no nucleation event takes place in time interval $dt'$, with $0 < t' < t$, in the disk of area $\pi R^2(t, t')$ given by

$$\pi R^2(t, t', h) = \pi[r_n^2(t, t') - h^2], \quad (A10)$$

where $r_n(t, t')$ is the nucleus growth law and $h \leq h_{max} = r_n(t, 0)$. The stochastic problem is therefore equivalent to a stochastic process of dots in 2D-space. For constant nucleation rate the equation for $P(t, h)$ becomes

$$P(t, h) = \exp\left[-I_p \int_0^{\bar{t}} \pi R^2(t, t', h) dt'\right]. \qquad (A11)$$

In eqn.A11 the extreme of integration, $\bar{t}(t)$, satisfies the equation $R(t, \bar{t}, h) = 0$. In the case of parabolic growth law, $r_n(t, t') = \sqrt{c(t - t')}$ (with constant c) eqn.A11 implies [6]

$$P(t, h) = \exp\left[-\pi I_p \int_0^{\bar{t}} [c(t - t') - h^2] dt'\right] = \exp\left[-\frac{\pi I_p c t^2}{2}\left(1 - \frac{h^2}{ct}\right)^2\right]. \qquad (A12)$$

The volume of the deposit, per unit area, is eventually computed through integration of eqn.A12 over $h$ up to the maximum height $h_{max} = \sqrt{ct}$ according to

$$V(t) = \int_0^{h_{max}} \left(1 - P(t, h)\right) dh = \frac{\sqrt{ct}}{2} \int_0^1 \frac{1}{\sqrt{\eta}} \left(1 - e^{-V_e(t)(1-\eta)^2}\right) d\eta, \qquad (8)$$

where $\eta = h^2/ct$ and $V_e(t) = \frac{\pi I_p c t^2}{2}$ is the phantom included extended surface. The normalized volume per unit area reads $W(t) = \frac{V(t)}{\sqrt{ct}}$.

---

[6] In eqn.A11, $\bar{t} = t - \frac{h^2}{c}$ and the integral reads $\int_{\frac{h^2}{c}}^t [cx - h^2] dx = \left[c\frac{x^2}{2} - h^2 x\right]_{\frac{h^2}{c}}^t = \frac{1}{2}ct^2\left(1 - \frac{h^2}{ct}\right)^2$.